\documentstyle[12pt]{article}
\title{\bf
COHERENT AND TRAJECTORY-COHERENT STATES OF A DAMPED HARMONIC OSCILLATOR}
\author{{\bf A.G. Karavayev and Yu.O. Budayev \ }\\
{\it Tomsk Polytechnic University , 634004 Tomsk, Russia }}
\date{}
\begin{document}
\maketitle
\begin{abstract}
\noindent
In this paper we construct the coherent and  trajectory-coherent states
of a damped harmonic oscillator. We investigate the properties of this states.
\end{abstract}
\begin{flushleft}
\subsection*{1.~Introduction.}
\end{flushleft}
        At present the coherent states are widely used to describe
many fields of theoretical physics.$^1$ The interest and activity
in coherent states were revived by the paper of Glauber,$^2$\ who
showed that the coherent states could be successfully used for
problem of quantum optics.Recently Nieto and Simmons$^3$\ have constructed
coherent states for particles in general potential,Hartley and Ray$^4$\ have
obtained coherent states for time-dependent harmonic oscillator on the basis of
Lewis-Risenfeld theory;Yeon,Um and George$^5$\ have constructed exact coherent
states for the damped harmonic oscillator.

Some time ago  Bagrov,Belov   and  Ternov$^6$\ have    constructed
approximate  (for  $  \hbar  \rightarrow  {  0}$  )  solutions  of the
Schr\"{o}dinger equation  for particles   in   general   potentials,  such
that  the coordinate and  momentum quantum - mechanical averages  were
exact  solutions    of   the  corresponding   classical    Hamiltonian
equations;   these  states   were  called  trajectory- coherent (TCS).
The  basis  of  this   construction  is  the  complex  WKB  method  by
V.P.Maslov.$^{7-9}$

    The  aim  of   this  work  is   to construct the  coherent (CS)
and trajectory-coherent (TCS) states of a damped harmonic oscillator
by using the Caldirola - Kanai Hamiltonian and the  complex  WKB  method.
It is shown that this states satisfy the Schr\"{o}dinger equation exactly.

\begin{flushleft}
\subsection*{2.~The construction of CS and TCS.}
\end{flushleft}
        Consider the Schr\"{o}dinger equation
\begin{equation}
    i\hbar\partial_t\Psi=\hat{H}\Psi,                          
\end{equation}
where the symbol of operator  $\hat{H}$  -  the  function  $H(x,p,t)$
is arbitrary real  and  analytical  function  of   coordinate  and
momentum.The  method  of  construction  TCS  in  this case has been
described  in  detail  in Ref.6,\ hence  we  shall illustrate some
moments only.  For constructing the  TCS of the Schr\"{o}dinger equation
it is necessary to solve the classical Hamiltonian system
\begin{equation}
   \dot{x}(t)=\partial_p{H(x,p,t)},~ \dot{p}(t)=-\partial_x{H(x,p,t)},
\end{equation}
and the system in variations (this is the linearization of the
Hamiltonian system in the neighbourhood of the trajectory $x(t),p(t)$)
\begin{eqnarray}
\dot{w}(t)=-H_{xp}(t)w(t)-H_{xx}(t)z(t),&   w(0)=b,   \\             
\dot{z}(t)=H_{pp}(t)w(t)+H_{px}(t)z(t), &   z(0)=1, \nonumber
\end{eqnarray}
where $H(x,p,t)$ is the classical Hamiltonian
\begin{eqnarray}
H_{xp}(t)= \partial_x\partial_p{H(x,p,t)} \mid_{x=x(t),p=p(t)},~
H_{px}(t)= \partial_p\partial_x{H(x,p,t)} \mid_{x=x(t),p=p(t)},\nonumber\\
H_{xx}(t)= \partial_{xx}^2 {H(x,p,t)} \mid_{x=x(t),p=p(t)},~
H_{pp}(t)= \partial_{pp}^2 {H(x,p,t)} \mid_{x=x(t),p=p(t)}, \nonumber
\end{eqnarray}
$b$ is complex number obeying the condition $ { \bf Im}b> 0$,~and
~$x(t),~p(t)$~are the solutions of system (2).

Consider the damped harmonic oscillator$^{5,10,11}$
\begin{equation}
  H(x,p,t)=\exp(-\gamma t)(2m)^{-1}p^2+\frac{1}{2}\exp(\gamma t)m
\omega_0^2 x^2.                                                 
\end{equation}
The Lagrangian and mechanical energy are given by$^5$
\begin{eqnarray}
L=\exp(\gamma t)(\frac {1}{2}m\dot{x}^2-\frac {1}{2}m\omega_0^2 x^2),~
E=\exp(-2\gamma t)(2m)^{-1}p^2+\frac{1}{2}m\omega_0^2 x^2.      
\end{eqnarray}
We first define the Cauchy problem for equation (1)
\begin{eqnarray}
\mid 0>\mid_{t=0}=\Psi_0(x,t,\hbar)\mid_{t=0}=
N\exp\{i\hbar^{-1}(p_0(x-x_0)+\frac{b}{2}
(x-x_0)^2)\},                                                   
\end{eqnarray}
where $x_0=x(t)\mid_{t=0},~p_0=p(t)\mid_{t=0}.$

    The function of WKB - solution type (TCS)$^6$
\begin{equation}
\mid 0>=\Psi_0(x,t,\hbar)=N\Phi(t)\exp\{i\hbar^{-1}S(x,t)\},     
\end{equation}
where $N=({\bf Im}b(\pi\hbar)^{-1})^{-1/4},~ \Phi(t)=(z(t))^{-1/2},$
\begin{eqnarray}
S(x,t)=\int\limits_{0}^{t}{\{\dot{x}(t)p(t)-H(x(t),p(t),t)\}dt}+ \nonumber
p(t)(x-x(t))+ \\+\frac{1}{2}w(t)z^{-1}(t)(x-x(t))^{2},\nonumber
\end{eqnarray}
and the phase $S(x,t)$ is the complex - valued function $({\bf Im}S> 0)$
is the approximate solution of the Cauchy problem (6) for the Schr\"{o}dinger
equation (1).
We should note that in the case of the quadratic systems,
for example, for Hamiltonian (4) the function (7) is the
exact solution of the equation (1).

Solving the differential equations (2),(3) we obtain
\begin{equation}
x(t)=\cosh\omega t\exp(-\frac{1}{2}\gamma t),~ p(t)=m(\omega
\sinh \omega t -\frac{1}{2}\gamma \cosh \omega t)\exp(\frac{1}{2}\gamma t),
\end{equation}                                                     
$$
w(t)=m\exp(\frac{1}{2}\gamma t)\{(\delta\omega-\frac{1}{2}\gamma)\cosh \omega
t+ (\omega-\frac{1}{2}\gamma\delta) \sinh \omega t\},
$$
$$
z(t)=\exp(-\frac{1}{2}\gamma t)\{\cosh\omega t+\delta \sinh \omega t\},
$$
$$
 \delta =(m\omega)^{-1}(b+\frac{1}{2}\gamma m),~ \omega^2=\frac{1}{4}\gamma^2
-\omega_0^2.
$$
It is easy to check that the function (7), where~$ x(t),p(t),w(t),z(t)$~
are defined by (8) satisfy the equation (1) exactly as for ~$
\omega^2>0$~as ~$\omega^2<0~~(\omega \rightarrow i\omega)$~.
\pagebreak

Further,we define annihilation operator~ $\hat{a}(t)$~ and creation operator
~$\hat{a}^+(t)$ as:$^6$
\begin{equation}
\hat{a}(t)=(2\hbar {\bf Im}b)^{-1/2} \{z(t)(\hat{p}-p(t))-w(t)(x-x(t))\},
\end{equation}                                           
$$
\hat{a}^+(t)=(2\hbar {\bf Im}b)^{-1/2} \{z^*(t)(\hat{p}-p(t))-w^*(t)(x-x(t))\}.
$$
It easy to check that the creation and annihilation operators satisfy
the usual Bose permutation rule
\begin{equation}
[\hat{a},\hat{a}^+]=1,~[\hat{a},\hat{a}]= [\hat{a}^+,\hat{a}^+]=0.  
\end{equation}
The complete orthonormal set of the trajectory-coherent states (TCS)
are defined as:$^6$
\begin{equation}
\mid n>=(n!)^{-1/2}(\hat{a}^+)^n\mid 0>.   
\end{equation}
It is not difficult to check the relations
\begin{equation}
 <n\mid m>=\delta_{n,m},~~~~\hat{a}\mid 0>=0,\\
\end{equation}
$$
 \hat{a}^+\mid n>=(n+1)^{1/2}\mid n+1>,~~~
\hat{a}\mid n>=(n)^{1/2}\mid n-1>. \nonumber         
$$

Using the usual methods$^1$ we obtain the expression for coherent
states (CS)
\begin{equation}
\mid \alpha >=\hat{A}(\alpha )\mid 0>,             
\end{equation}
where $\hat{A}(\alpha )$ is the unitary operator
\begin{equation}
\hat{A}(\alpha )=\exp (\alpha \hat{a}^+-\alpha ^* \hat{a}),   
\end{equation}
~$\alpha $~ is the complex number;~ $\hat{a}^+, \hat{a}$~ are defined
by (9).

Besides, at is follows from (12)-(14), the functions~ $\mid \alpha >$~ are
eigenstates of the operator~ $\hat{a}$~ with eigenvalue~ $\alpha $,\ i.e.
\begin{equation}
\hat{a} \mid \alpha >=\alpha \mid \alpha >.               
\end{equation}
\begin{flushleft}
\subsection*{3.\ Quantum - mechanical averages
and uncertainty \\ \ \ \ relations.}
\end{flushleft}
Further we shall find the expressions for quantum - mechanical averages
$$
<\hat{x}>_{\scriptscriptstyle TCS},\
<\hat{x}^2>_{\scriptscriptstyle TCS},\
<\hat{p}>_{\scriptscriptstyle TCS},\
<\hat{p}^2>_{\scriptscriptstyle TCS},\
<\hat{x}>_{\scriptscriptstyle CS},\
$$
$$
<\hat{x}^2>_{\scriptscriptstyle CS},
<\hat{p}>_{\scriptscriptstyle CS},\
<\hat{p}^2>_{\scriptscriptstyle CS},\
<\hat{E}>_{\scriptscriptstyle TCS},\
<\hat{E}>_{\scriptscriptstyle CS},\
$$
where, for example,
$$
<\hat{x}>_{\scriptscriptstyle TCS}=<n\mid \hat{x} \mid n>,~
<\hat{x}>_{\scriptscriptstyle CS}=<\alpha \mid \hat{x} \mid \alpha>.
$$

Further,we present the relations expressing the coordinate
and momentum operators in terms of the operators ~ $\hat{a}^+, \hat{a}$~
\begin{equation}                   
\hat{x}=x(t)-i(2{\bf
Im}b(\hbar)^{-1})^{-1/2}\{z(t)\hat{a}^+-z^*(t)\hat{a}\},
\end{equation}
$$
\hat{p}=p(t)-i(2{\bf
Im}b(\hbar)^{-1})^{-1/2}\{w(t)\hat{a}^+-w^*(t)\hat{a}\}.
$$
Using (5),(9)-(16), we obtain for quantum - mechanical averages:
$$
<\hat{x}>_{\scriptscriptstyle TCS}=x(t),~
<\hat{p}>_{\scriptscriptstyle TCS}=p(t),
$$
\begin{equation}                   
<\hat{x}>_{\scriptscriptstyle CS}=x(t)-i(2{\bf
Im}b(\hbar)^{-1})^{-1/2}\{ \alpha^* z(t)- \alpha z^*(t)\},
\end{equation}
$$
<\hat{p}>_{\scriptscriptstyle CS}=p(t)-i(2{\bf
Im}b(\hbar)^{-1})^{-1/2}\{\alpha^* w(t)-\alpha w^*(t)\},
$$
$$
<\hat{x}^2>_{\scriptscriptstyle TCS}=
x^2(t)+\frac{\hbar}{{\bf Im}b}(n+\frac{1}{2})|{z(t)}|^2,
$$
$$
<\hat{p}^2>_{\scriptscriptstyle TCS}=
p^2(t)+\frac{\hbar}{{\bf Im}b}(n+\frac{1}{2})|{w(t)}|^2,
$$
$$
<\hat{x}^2>_{\scriptscriptstyle CS}=
<x>_{\scriptscriptstyle CS}^2+\frac{\hbar}{2{\bf Im}b}|{z(t)}|^2,
$$
$$
<\hat{p}^2>_{\scriptscriptstyle CS}=
<p>_{\scriptscriptstyle CS}^2+\frac{\hbar}{2{\bf Im}b}|{w(t)}|^2.
$$
\begin{eqnarray*}
<\hat{E}>_{\scriptscriptstyle TCS}=
\exp(-2\gamma t)(2m)^{-1}p^2(t)
+\frac{1}{2}m\omega_0^2 x^2(t)+ \\
\frac{\hbar}{{\bf Im}b}(n+\frac{1}{2})
\{\exp(-2\gamma t)(2m)^{-1}|{w(t)}|^2
+\frac{1}{2}m\omega_0^2|{z(t)}|^2\},
\end{eqnarray*}
\begin{eqnarray*}
<\hat{E}>_{\scriptscriptstyle CS}=
\exp(-2\gamma t)(2m)^{-1}<p>_{\scriptscriptstyle CS}^2
+\frac{1}{2}m\omega_0^2<x>_{\scriptscriptstyle CS}^2+ \\
\frac{\hbar}{2{\bf Im}b}
\{\exp(-2\gamma t)(2m)^{-1}|{w(t)}|^2
+\frac{1}{2}m\omega_0^2|{z(t)}|^2\}.
\end{eqnarray*}
Now, by calculating the uncertainty in\ $x$\   and\ $\hat{p}$\
in the TCS and CS one finds:
\begin{eqnarray}
(\Delta\hat{x})_{\scriptscriptstyle TCS}^2=
<(\hat{x}-<\hat{x}>_{\scriptscriptstyle TCS})^2>_{\scriptscriptstyle TCS}
=\hbar(n+\frac{1}{2})\frac{|{z(t)}|^2}{{\bf Im}b}, \\  
(\Delta\hat{p})_{\scriptscriptstyle TCS}^2=
<(\hat{p}-<\hat{p}>_{\scriptscriptstyle TCS})^2>_{\scriptscriptstyle TCS}
=\hbar(n+\frac{1}{2})\frac{|{w(t)}|^2}{{\bf Im}b}, \nonumber
\end{eqnarray}
\begin{eqnarray}
(\Delta\hat{x})_{\scriptscriptstyle CS}^2=
<(\hat{x}-<\hat{x}>_{\scriptscriptstyle CS})^2>_{\scriptscriptstyle CS}
= \frac{\hbar}{2} \frac{|{z(t)}|^2}{{\bf Im}b}, \\  
(\Delta\hat{p})_{\scriptscriptstyle CS}^2=
<(\hat{p}-<\hat{p}>_{\scriptscriptstyle CS})^2>_{\scriptscriptstyle CS}
= \frac{\hbar}{2} \frac{|{w(t)}|^2}{{\bf Im}b}. \nonumber
\end{eqnarray}
So,\ the Heisenberg's uncertainty relations is expressed as
\begin{equation}   
(\Delta\hat{x})_{\scriptscriptstyle TCS}^2
(\Delta\hat{p})_{\scriptscriptstyle TCS}^2=
\hbar^2 (n+\frac{1}{2})^2 \frac{| w(t)z(t) | ^2}{({\bf Im}b)^2},
\end{equation}
\begin{equation}                   
(\Delta\hat{x})_{\scriptscriptstyle CS}^2
(\Delta\hat{p})_{\scriptscriptstyle CS}^2=
\frac{\hbar^2}{4} \frac{| w(t)z(t) | ^2}{({\bf Im}b)^2}.
\end{equation}

For the damped harmonic oscillator (4) we choose\ ${\bf Re}b=0$\
(it is necessary for minimization of the uncertainty relations in
the initial instant of time),\ ${\bf Im}b=\mu m | \omega | $\ ,
where\ $\mu >0$\ shows the initial uncertainty of coordinate;\ and we
denote\ $\theta = \gamma /{2\omega}$.

By using (8),(16)-(21), in the case~\
$\omega^2=\omega_0^2-\frac{1}{4}\gamma^2>0$~ for\ \ $
(\Delta\hat{x})_ {\scriptscriptstyle TCS}^2,\\ (\Delta\hat{p})_
{\scriptscriptstyle TCS}^2 ,\
(\Delta\hat{x})_
{\scriptscriptstyle CS}^2,\ (\Delta\hat{p})_
{\scriptscriptstyle CS}^2$\ \ we obtain
\begin{eqnarray*}
(\Delta\hat{x})_{\scriptscriptstyle TCS}^2
=\hbar(n+\frac{1}{2})\exp(-\gamma t) (\mu m \omega)^{-1}
\{1+\sin^2\omega t(\theta^2+\\+\mu^2-1)+\theta \sin2\omega t\},
 \nonumber
\end{eqnarray*}
\begin{eqnarray*}
(\Delta\hat{p})_{\scriptscriptstyle TCS}^2
=\hbar(n+\frac{1}{2})\exp(\gamma t)\mu m
\omega\{1+\frac{1}{\mu^2}\sin^2\omega
t(2\theta^2+\\+\theta^4+1+\mu^2\theta^2-\mu^2)-\theta\sin2\omega t\},
\nonumber
\end{eqnarray*}
\begin{eqnarray*}
(\Delta\hat{x})_{\scriptscriptstyle CS}^2
=\frac{\hbar}{2}\exp(-\gamma t) (\mu m \omega)^{-1}
\{1+\sin^2\omega t(\theta^2+\\+\mu^2-1)+\theta \sin2\omega t\},
 \nonumber
\end{eqnarray*}
\begin{eqnarray*}
(\Delta\hat{p})_{\scriptscriptstyle CS}^2
=\frac{\hbar}{2}\exp(\gamma t)\mu m
\omega\{1+\frac{1}{\mu^2}\sin^2\omega
t(2\theta^2+\\+\theta^4+1+\mu^2\theta^2-\mu^2)-\theta\sin2\omega t\},
\nonumber
\end{eqnarray*}
and thus the uncertainty relations becomes
\begin{equation}
(\Delta\hat{x})_{\scriptscriptstyle TCS}^2
(\Delta\hat{p})_{\scriptscriptstyle TCS}^2
=\hbar^2(n+\frac{1}{2})^2(1+g(t)), 
\end{equation}
$$
(\Delta\hat{x})_{\scriptscriptstyle CS}^2
(\Delta\hat{p})_{\scriptscriptstyle CS}^2
=\frac{\hbar^2}{4}(1+g(t)),
$$
where
\begin{equation}
g(t)=\{\frac{\theta}{\mu}(\theta^2+\mu^2+1)\sin^2\omega t+
\frac{1}{2\mu}(\theta^2-\mu^2+1)\sin2\omega t\}^2.      
\end{equation}
We should note,that in the special case\ $\mu =1$\  the
formula (23) coincides with formula (14) from Ref.5.

The function (23) is equal to zero,and,therefore, the
minimization of the uncertainty relations has place in the  instants of
time \begin{eqnarray*} t_{1k}=\frac{\pi
k}{\omega},~~~t_{2k}=\frac{1}{\omega} \arctan
\frac{\mu^2-\theta^2-1}{\theta(\mu^2+\theta^2+1)}+\frac{\pi k}{\omega};\ ~
k=0,1,2,\ldots .
\end{eqnarray*}
The preceding equation can be solved for the parameter\ $\mu >0$\ in
the case
\begin{equation}
|\theta \tan \omega t|<1.              
\end{equation}
and,therefore,by choosing parameter\ $\mu $\ we can obtain the
minimization for any instant of time\ $t$\ obeying the condition (24).

In the case\  $\omega^2=\frac{1}{4}\gamma^2-\omega_0^2>0$\ we obtain
\begin{eqnarray*}
(\Delta\hat{x})_{\scriptscriptstyle TCS}^2
=\hbar(n+\frac{1}{2})\exp(-\gamma t) (\mu m \omega)^{-1}
\{1+\sinh^2\omega t(\theta^2+\\+\mu^2+1)+\theta \sinh2\omega t\},
 \nonumber
\end{eqnarray*}
\begin{eqnarray*}
(\Delta\hat{p})_{\scriptscriptstyle TCS}^2
=\hbar(n+\frac{1}{2})\exp(\gamma t)\mu m
\omega\{1+\frac{1}{\mu^2}\sinh^2\omega
t(1-2\theta^2+\\+\theta^4+\mu^2\theta^2+\mu^2)-\theta\sinh2\omega t\},
\nonumber
\end{eqnarray*}
\begin{eqnarray*}
(\Delta\hat{x})_{\scriptscriptstyle CS}^2
=\frac{\hbar}{2}\exp(-\gamma t) (\mu m \omega)^{-1}
\{1+\sinh^2\omega t(\theta^2+\\+\mu^2+1)+\theta \sinh2\omega t\},
 \nonumber
\end{eqnarray*}
\begin{eqnarray*}
(\Delta\hat{p})_{\scriptscriptstyle CS}^2
=\frac{\hbar}{2}\exp(\gamma t)\mu m
\omega\{1+\frac{1}{\mu^2}\sinh^2\omega
t(1-2\theta^2+\\+\theta^4+\mu^2\theta^2+\mu^2)-\theta\sinh2\omega t\},
\nonumber
\end{eqnarray*}
and for the uncertainty relations we have
$$
(\Delta\hat{x})_{\scriptscriptstyle TCS}^2
(\Delta\hat{p})_{\scriptscriptstyle TCS}^2
=\hbar^2(n+\frac{1}{2})^2(1+g(t)),
$$
$$
(\Delta\hat{x})_{\scriptscriptstyle CS}^2
(\Delta\hat{p})_{\scriptscriptstyle CS}^2
=\frac{\hbar^2}{4}(1+g(t)),
$$
where
\begin{equation}
g(t)=\{\frac{\theta}{\mu}(\theta^2+\mu^2-1)\sinh^2\omega t+
\frac{1}{2\mu}(\theta^2-\mu^2-1)\sinh2\omega t\}^2.    
\end{equation}
Now the function\ $g(t)$\ (25) is equal to zero only for
\begin{eqnarray*}
~~t_{01}=0~~,~~t_{02}=\frac{1}{\omega}{\rm
arctanh}\frac{\mu^2-\theta^2+1}{\theta(\mu^2+ \theta^2-1)}.
\end{eqnarray*}
This equation can be solved for parameter \ $\mu $\ in the
case
\begin{equation}
|\theta \tanh \omega t|<1,                                   
\end{equation}
and we can obtain the minimization for any instant of time\ $t$\
obeying the condition (26).

\begin{flushleft}
\subsection*{Acknowledgments}
\end{flushleft}
We thank Prof.\ A.Yu.Trifonov and Prof.\ D.V.Boltovsky for
useful discussions.
\pagebreak

\end{document}